\documentclass[prd,nofootinbib,twocolumn,superscriptaddress,showpacs]{revtex4}
\usepackage{graphicx}
\usepackage{bm}
\usepackage{natbib}
\usepackage{amsfonts} 
\usepackage{amssymb}
\usepackage{bbm}
\usepackage{epsfig}
 \usepackage{epstopdf}
 \usepackage{amsmath,amssymb}
\usepackage{graphics}
\usepackage{subfigure}
\usepackage{colordvi}
\usepackage{color}

\begin{document}

\newcommand{\be}{\begin{equation}}
\newcommand{\ee}{\end{equation}}
\newcommand{\bea}{\begin{eqnarray}}
\newcommand{\eea}{\end{eqnarray}}

\newcommand{\kp}{\kappa}
\newcommand{\Om}{\Omega}
\newcommand{\de}{\Delta}
\newcommand{\eps}{\epsilon}
\newcommand{\g}{\gamma}

\newcommand{\tot}{\mathrm{tot}}
\newcommand{\ini}{\mathrm{in}}
\newcommand{\eff}{\mathrm{eff}}
\newcommand{\nr}{\mathrm{NR}}
\newcommand{\e}{\mathrm{end}}
\newcommand{\h}{\mathcal{H}}
\newcommand{\Hyd}{\mathrm{H}}
\newcommand{\He}{\mathrm{He}}
\newcommand{\eV}{\mathrm{eV}}
\newcommand{\keV}{\mathrm{keV}}
\newcommand{\GeV}{\mathrm{GeV}}
\newcommand{\lya}{Ly$\alpha \ $}
\newcommand{\iMpc}{\mbox{ Mpc$^{-1}$}}
\newcommand{\cms}{cm$^3$s$^{-1}$}
\newcommand{\sv}{\langle \sigma_{\mathrm{A}} v \rangle}
\newcommand{\ud}{\mbox{d}}
\newcommand{\fesc}{f_{\mathrm{esc}}}
\newcommand{\fa}{f_{\alpha}}
\newcommand{\fx}{f_{X}}
\newcommand{\Nion}{N_{\mathrm{i}}}
\newcommand{\sigv}{\langle \sigma_A v \rangle }
\newcommand{\ic}{\mathrm{IC}}
\newcommand{\prompt}{\mathrm{prompt}}
\newcommand{\Mmin}{M_{\mathrm{min}}}
\newcommand{\nA}{n_{\mathrm{A}}}
\newcommand{\nH}{n_{\mathrm{H}}}
\newcommand{\nHe}{n_{\mathrm{He}}}
\newcommand{\Th}{\theta_{13}}
\newcommand{\nua}{\nu_{\alpha}}
\newcommand{\fin}{\mathrm{fin}}

\newcommand\pp{\,\,\,.}
\newcommand\vv{\,\,\,,}

\newcommand{\apjl}{Astrophys. J. Lett.}
\newcommand{\apjs}{Astrophys. J. Suppl. Ser.}
\newcommand{\aap}{Astron. \& Astrophys.}
\newcommand{\aj}{Astron. J.}
\newcommand{\araa}{Ann. Rev. Astron. Astrophys. } 
\newcommand{\mnras}{Mon. Not. R. Astron. Soc.}
\newcommand{\physrep}{Phys. Rept.}
\newcommand{\jcap}{JCAP}

\begin{minipage}[t]{6.9in}
\hfill{\tt CERN-PH-TH-2012-089, IFIC/12-28, LAPTH-018/12}
\end{minipage}

\title{Cosmological lepton asymmetry with a nonzero mixing angle $\theta_{13}$}

\author{Emanuele Castorina}
\affiliation{SISSA, Via Bonomea 265, 34136, Trieste, Italy}

\author{Urbano Fran\c{c}a}
\affiliation{Instituto de F\'{\i}sica Corpuscular  (CSIC-Universitat de Val\`{e}ncia), 
Apdo.\ 22085, 46071 Valencia, Spain}

\author{Massimiliano Lattanzi}
\affiliation{Dipartimento di Fisica G. Occhialini, Universit{\`a} Milano-Bicocca and INFN,
Sezione di Milano-Bicocca, Piazza della Scienza 3, I-20126 Milano, Italy}

\author{Julien Lesgourgues}
\affiliation{CERN, Theory Division, CH-1211 Geneva 23, Switzerland}
\affiliation{Institut de Th{\'e}orie des Ph{\'e}nom{\`e}nes Physiques, EPFL, CH-1015 Lausanne, Switzerland}
\affiliation{LAPTH (CNRS-Universit{\'e} de Savoie), B.P. 110, F-74941 Annecy-le-Vieux Cedex, France}

\author{Gianpiero Mangano}
\affiliation{INFN, Sezione di Napoli, Complesso Univ. Monte S. Angelo, Via Cintia, I-80126 Napoli, Italy}

\author{Alessandro Melchiorri}
\affiliation{Physics Department and INFN, Universit{\`a} di Roma “La Sapienza”, Ple Aldo Moro 2, 00185, Rome, Italy}

\author{Sergio Pastor}
\affiliation{Instituto de F\'{\i}sica Corpuscular  (CSIC-Universitat de Val\`{e}ncia), 
Apdo.\ 22085, 46071 Valencia, Spain}

\date{\today}

\begin{abstract}
While the baryon asymmetry of the Universe is nowadays well measured 
by cosmological observations, the bounds on the lepton 
asymmetry in the form of neutrinos are still significantly weaker. 
We place limits on the relic 
neutrino asymmetries using some of the latest cosmological data,
taking into account the effect of flavor oscillations. 
We present our results for two different values of the 
neutrino mixing angle $\theta_{13}$,  and show that for large 
$\theta_{13}$ the limits on the total neutrino asymmetry become more 
stringent, diluting even large initial flavor asymmetries. In 
particular, we find that the present bounds are still dominated 
by the limits coming from Big Bang Nucleosynthesis, while the 
limits on the total neutrino mass from cosmological data 
are essentially independent of $\theta_{13}$. Finally, 
we perform a forecast for COrE, taken as an example 
of a future CMB experiment, and find that it could improve the 
limits on the total lepton asymmetry approximately by up to a factor 6.6.
\end{abstract}

%

%

\maketitle


\section{Introduction} \label{sec:intro}

Quantifying the asymmetry between matter and antimatter of the Universe
is crucial for understanding some of the particle physics processes
that might have taken place in the early Universe, at energies much
larger than the ones that can be reached currently in particle accelerators. 
Probes of the anisotropies of the cosmic microwave 
background (CMB) together with other
cosmological observations have 
measured the  cosmological baryon asymmetry $\eta_b$ to the percent level 
thanks to very precise measurements of the baryon density \cite{Komatsu11}. 
For the lepton asymmetries, while they are expected to be of the same order 
of the baryonic one due to sphaleron effects that equilibrate both asymmetries, 
it could be the case that other physical processes lead instead
to leptonic asymmetries much larger than $\eta_b$ (see, {\it e.g.},
 \cite{Casas99,March99,McDonald99}), with consequences for the early Universe 
phase transitions \cite{Schwarz09}, cosmological magnetic fields  \cite{Semikoz09},
and the dark matter relic density \cite{Shi99,Laine08,Stuke11}. Neutrino asymmetries are also 
bound to be nonzero in the presence of neutrino isocurvature perturbations, like those
generated by curvaton decay \cite{Lyth:2002my,Gordon:2003hw,DiValentino12}. 
Those large neutrino asymmetries could have been imprinted in the
cosmological data \cite{Lesgourgues99,Lesgourgues06}, and although the limits
on such asymmetries have been improving over the last years, current constraints 
are still many orders of magnitude weaker than the baryonic measurement.
                                   
On the other hand, thanks to the neutrino oscillations the initial primordial flavor asymmetries are redistributed among the active neutrinos before the onset of Big Bang Nucleosynthesis (BBN) \cite{Dolgov02,Wong02,Abazajian02}, which makes the
knowledge of the oscillation parameters important for correctly interpreting the
limits on such asymmetries. Nowadays all of those parameters are 
accurately measured (see {\it e.g.} \cite{Fogli11,Schwetz11}), with the exception of the mixing angle $\theta_{13}$ that
only recently started to be significantly constrained.
In fact, several neutrino experiments over the last year gave indications
of nonzero values for $\sin^2 \theta_{13}$ \cite{t2k11, minos11, chooz11}, and recently
the Daya Bay reactor experiment claimed a measurement of $\sin^2 (2\theta_{13}) = 0.092 \pm 0.016$(stat.)$\pm0.005$(syst.)
at 68\% C.L.~\cite{dayabay12}, excluding a zero value for $\theta_{13}$ with high significance. The same finding has been also reported by the RENO Collaboration \cite{collaboration:2012nd}, $\sin^2 (2 \theta_{13}) = 0.113 \pm 0.013$(stat.)$\pm0.019$(syst.) (68\% C.L.).

Finally, yet another important piece of information for reconstructing the
neutrino asymmetries in the Universe is the measured value of the 
relativistic degrees of freedom in the 
early universe, quantified in the so-called effective number of neutrinos,
$N_{\eff}$. In the case of the three active neutrino flavors 
with zero asymmetries and a standard thermal history, its
value is the well-known $N_{\eff} \simeq 3.046$ \cite{Mangano05}, but the presence
of neutrino asymmetries can increase that number 
while still satisfying the BBN constraints \cite{Pastor09}. 
Interestingly enough, recent CMB data has 
consistently given indications of $N_{\eff}$ higher than the standard value: recently
the Atacama Cosmology Telescope (ACT) \cite{act11} and 
the South Pole Telescope (SPT)  \cite{spt11, spt12} have found evidence
for $N_{\eff} > 3.046$ at 95$\%$ C.~L., making the case for 
extra relativistic degrees  of freedom stronger (see also \cite{Archidiacono:2011gq}).
It should however be kept in mind that other physical processes, 
like {\it e.g.} the contribution from the energy density of sterile neutrinos \cite{Giusarma11,Hamann11} or of
gravitational waves \cite{Sendra2012}, could also lead to a larger value for $N_{\eff}$.

\begin{table*}
\caption{Cosmological and neutrino parameters.}
\label{tab:parameters}
\setlength{\tabcolsep}{12pt}
\begin{tabular}{lllc}
\hline \hline Type & Symbol & Meaning & Uniform Prior \\ \hline
Primary & $\Omega_b h^2$ & Baryon density & $(0.005, 0.1)$\\ 
Cosmological & $\Omega_{dm} h^2$ & Dark matter density\footnote{Also includes neutrinos.}& $(0.01, 0.99)$ \\ 
Parameters & $\tau$ & Optical depth to reionization & $ (0.01, 0.8)$ \\ 
$ \ $ & $100 \theta_s$ & Angular scale of the sound 
horizon at the last scattering & $(0.5, 10)$ \\ 
$ \ $ & $n_s$ & Scalar index of the power spectrum &$(0.5, 1.5)$ \\ 
$ \ $ & $\log\left[ 10^{10} A_{s} \right]$ & Scalar amplitude of the power spectrum \footnote{at the pivot wavenumber $k_0=0.05$ Mpc$^{-1}$.} & $(2.7, 4)$ \\ 
\hline 
Neutrino & $m_1$(eV) & Mass of the lightest neutrino  \footnote{We assume here normal hierarchy.} & $(0,  1)$ \\ 
Parameters & $\eta_{\nu}$ & Total asymmetry at $T=10$ MeV & $(-0.8, 0.8)$  \\
$\ $ & $\eta_{\nu_e}^{\ini}$ & Initial electron neutrino asymmetry at $T=10$ MeV& $(-1.2, 1.2)$  \\ 
\hline
Derived & $h$ & Reduced Hubble constant\footnote{$H_0=100 h$ km s$^{-1}$ Mpc$^{-1}$.} & - \\ 
Parameters & $\Delta N_{\eff}$ & Enhancement to the standard effective number of neutrinos\footnote{$N_{\eff} = 3.046$.}& -  \\
\hline
\end{tabular}
\setlength{\tabcolsep}{6pt}
\end{table*}

Some recent papers have analyzed 
the impact of neutrino asymmetries with oscillations on BBN \cite{Pastor09, Mangano11, Mangano12}, mainly
because  data on light element abundances dominate the current limits 
on the asymmetries.
Some studies using CMB data can be found  in the literature (see for instance 
\cite{Lattanzi:2005qq,Popa:2008tb,Shiraishi09} for limits on the 
degeneracy parameters $\xi_{\nu}$ using the WMAP data and \cite{Hamann07} for the
effect of the primordial Helium fraction in a Planck forecast), 
but our
paper improves on that in two directions. First, we used for our analysis
the neutrino spectra in the presence of asymmetries after taking into 
account the effect of flavor oscillations. 
Second, we checked the robustness of our results 
comparing the analysis of CMB and BBN data with
a more complete set of cosmological data, including in 
particular supernovae Ia (SNIa) data \cite{Kessler:2009ys}, the measurement of
the  Hubble constant from the Hubble Space Telescope (HST) \cite{Riess:2009pu}, 
and the Sloan Digital Sky Survey (SDSS) data on the matter power spectrum\cite{Reid:2009xm}. 
While current CMB measurements 
and the other datasets
are not expected to improve significantly the constraints on the
asymmetries, they constrain the sum
of the neutrino masses, giving a more robust and general picture 
of the cosmological
parameters. 

Our goals in this work 
is twofold: first, we constrain the
neutrino asymmetries and the sum of neutrino masses
for both zero and nonzero values of $\Th$ 
using some of the latest cosmological data to obtain an
updated and clear idea of the limits on them using current data; second,
we perform a forecast of the constraints that could be achievable with future CMB
experiments, taking as an example the proposed mission 
COrE\footnote{http://www.core-mission.org} \cite{core11}.
Given that current constraints are basically dominated by the
BBN constraints, we use our forecast to answer the more
general question of whether future CMB experiments can be 
competitive with the BBN bounds.

This paper is organized as follows. Initially, we 
briefly review in Sec.~\ref{sec:eqn} 
the dynamics of the neutrino asymmetries prior
to the BBN epoch. With those tools in hand,
we proceed to study in Sec.~\ref{sec:theta13} the impact on 
cosmological observables of
the neutrino asymmetries for two values of the mixing angle 
$\theta_{13}$ using current cosmological data. We then step towards the 
future and describe in Sec.~\ref{sec:forecast} our forecast
for the experiment COrE, where we study the potential of
the future data from lensing of CMB anisotropies 
to constrain some of the cosmological parameters (in particular, 
neutrino asymmetries and the sum of the neutrino masses) with great precision.
Finally, in Sec.~\ref{sec:conc} we draw our conclusions.


\section{Evolution of cosmological neutrinos with flavor asymmetries} \label{sec:eqn}

The dynamics of the neutrino distribution functions in the presence of
flavor asymmetries and neutrino oscillations in the early Universe has been discussed in
detail in the literature \cite{Pastor09, Mangano11, Mangano12}, and
here we will only briefly review its main features and its consequences for the late
cosmology.

We assume that flavor neutrino asymmetries,
$\eta_{\nua}$,
were produced in the early Universe. At large temperatures
frequent weak interactions keep neutrinos in equilibrium thus,
their energy spectrum follows a Fermi-Dirac distribution
with a chemical potential $\mu_{\nua}$ for
each neutrino flavor.
If $\xi_{\alpha} \equiv \mu_{\nua}/T$ is the
degeneracy parameter, the asymmetry is given by
\be \label{eq:eta}
\eta_{\nua}  \equiv \frac{n_{\nua} - n_{\bar{\nu}_\alpha}}{n_{\gamma}}=
\frac{1}{12 \zeta(3)} \left[ \pi^2 \xi_{\alpha} + \xi_{\alpha}^3\right] \ .
\ee
Here $n_{\nua}$ ($n_{\bar{\nu}_\alpha}$) denotes the
neutrino (antineutrino) number density, $n_{\gamma}$ is the photon number density,
and $\zeta(3)=1.20206$.

As usual, we will write the radiation energy density of the Universe in terms of
the parameter $N_{\eff}$, the effective number of neutrinos, as
\be
\rho_r = \rho_\gamma \left[ 1+ \frac{7}{8} \left(\frac{4}{11} \right)^{4/3} N_{{\rm eff}}\right] \ ,
\label{neff}
\ee
with $N_{{\rm eff}}=3.046$ the value in the standard case with zero asymmetries and
no extra relativistic degrees of freedom \cite{Mangano05}.
Assuming that equilibrium holds for the neutrino
distribution functions, the presence of flavor asymmetries leads to an
enhancement
\be
\Delta N_{\eff} =  \frac{15}{7}  \sum_{\alpha = e, \mu, \tau}\left[2
\left( \frac{\xi_{\alpha}}{\pi} \right)^2 +
\left( \frac{\xi_{\alpha}}{\pi} \right)^4\right] \ .
\label{deltan_xinu}
\ee
Note that a neutrino degeneracy
parameter of order $\xi_{\alpha}\gtrsim 0.3$
is needed in order to have a value of $\Delta N_{\eff}$
at least at the same level of the effect of non-thermal
distortions discussed in \cite{Mangano05}. This corresponds to
$\eta_{\nua}\sim{\cal O}(0.1)$. On the other hand,
the primordial abundance of $^4$He depends on the presence of an electron neutrino
asymmetry and sets  a stringent BBN bound on $\eta_{\nu_e}$ which does not apply to the other flavors,
leaving a total neutrino asymmetry of order unity unconstrained
\citep{Kang:1991xa,Hansen:2001hi}. However, this conclusion relies on the absence
of effective neutrino oscillations that would modify the distribution of the
asymmetries among the different flavors before BBN.

The evolution of the neutrino asymmetries in the epoch before BBN
with three-flavor neutrino oscillations is found by solving the equations of motion for
$3\times 3$ density matrices of the flavor neutrinos as described
in \cite{Sigl93, McKellar94}, including time-dependent vacuum and matter terms, both from
background $e^\pm$ and neutrinos, as well as
the collision integrals from neutrino weak interactions. This was done under certain approximations
in refs.\ \cite{Dolgov02,Wong02,Abazajian02}, where it was shown that neutrino
oscillations are indeed effective before the onset of BBN. Therefore, the total lepton
asymmetry is redistributed among the neutrino flavors and the BBN bound on $\eta_{\nu_e}$
can be translated into a limit on $\eta_\nu=\eta_{\nu_e}+\eta_{\nu_\mu}+\eta_{\nu_\tau}$,
unchanged by oscillations and constant until electron-positron annihilations, when
it decreases due to the increase in the photon number density.

The temperature at which flavor oscillations
become effective is important not only to establish $\eta_{\nu_e}$
at the onset of BBN, but also to determine whether weak interactions
with $e^+e^-$ can still keep neutrinos in good thermal contact with the primeval
plasma. Oscillations redistribute the asymmetries among the flavors, but
only if they occur early enough interactions would  preserve Fermi-Dirac spectra
for neutrinos, in such a way that the degeneracies $\xi_\alpha$ are
well defined for each $\eta_{\nu_\alpha}$ and the relation in Eq.\
(\ref{deltan_xinu}) remains valid. This the case of early conversions
of muon and tau neutrinos, since
oscillations and collisions rapidly equilibrate their
asymmetries at $T\simeq 15$ MeV \cite{Dolgov02}. Therefore one can assume the initial
values $\eta_{\nu_\mu}^{\rm in}=\eta_{\nu_\tau}^{\rm in}\equiv\eta_{\nu_x}^{\rm in}$, leaving
as free parameters $\eta_{\nu_e}^{\rm in}$ and the total
asymmetry $\eta_\nu=\eta_{\nu_e}^{\rm in}+2\eta_{\nu_x}^{\rm in}$.

\begin{figure*}[htpb]
\includegraphics[scale=0.5]{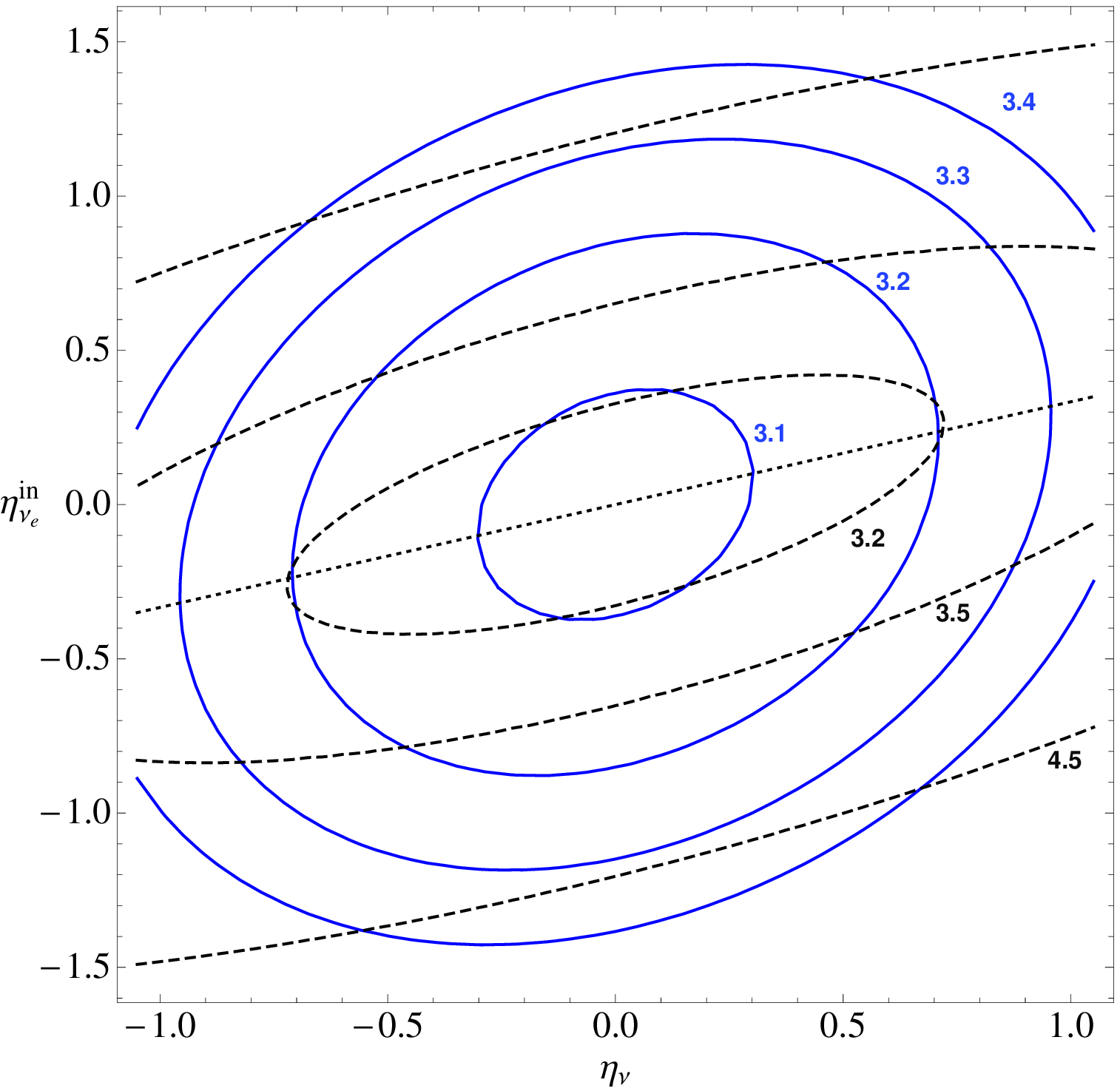}
\includegraphics[scale=0.5]{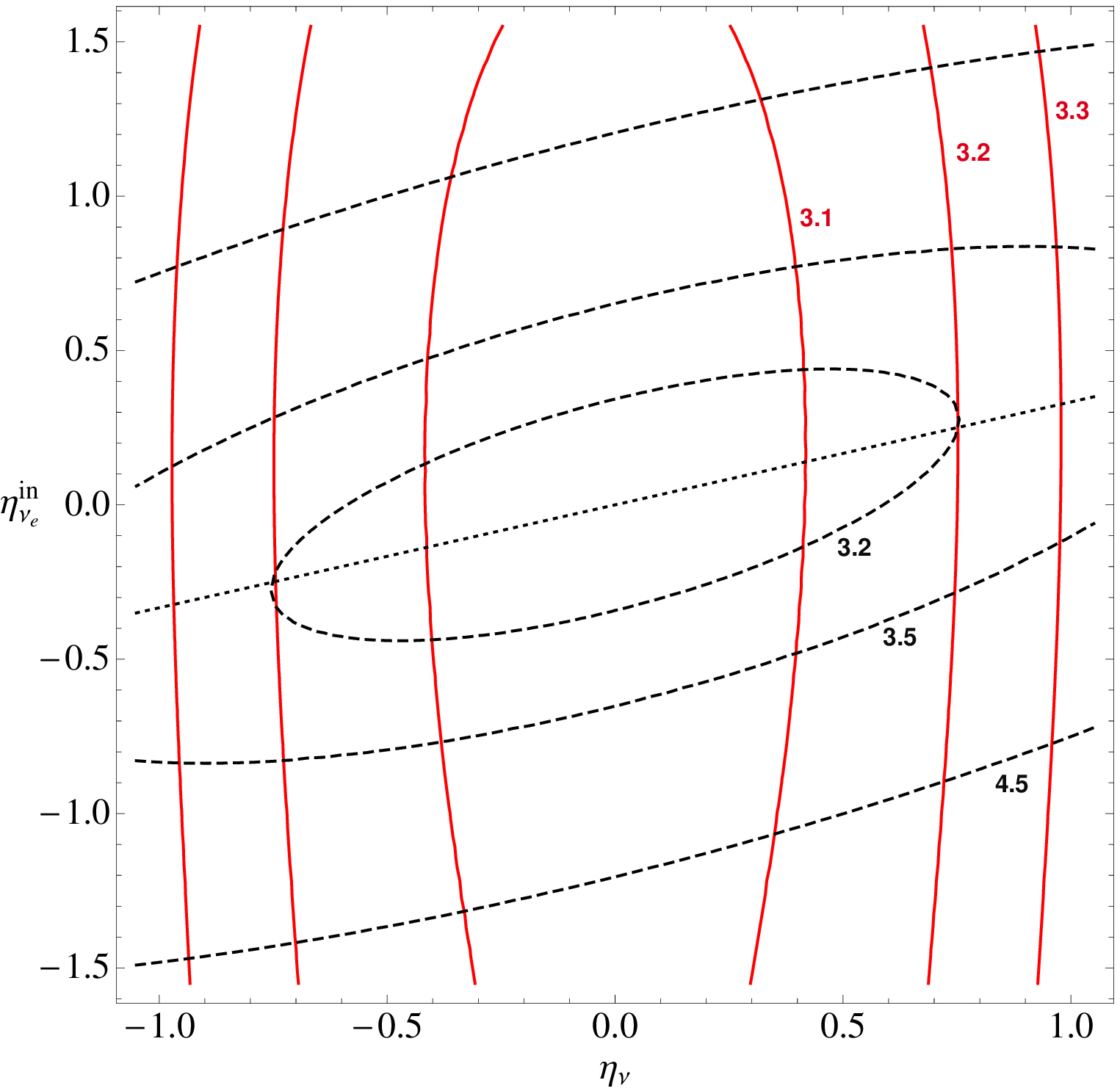}
\caption{Final contribution of neutrinos with 
primordial asymmetries to the radiation energy density.
The isocontours of $N_{\eff}$ on the  plane $\eta_{\nu_e}^{\ini}$ vs.\ $\eta_{\nu}$, including 
flavor oscillations, are shown for two values
of $\sin^2\theta_{13}$: $0$ (blue solid curves, left panel) and 
$0.04$ (red solid curves, right panel) and compared to the case with zero mixing (dashed curves).
 The dotted line corresponds to $\eta_{\nu}=\eta_{\nu_x}$ ($x = \mu, \tau$), where one 
expects oscillations to have negligible effects.}
\label{fig:asym}
\end{figure*}

If the initial values of the flavor asymmetries $\eta_{\nu_e}^{\rm in}$ and $\eta_{\nu_x}^{\rm in}$
have opposite signs, neutrino conversions will tend to reduce the asymmetries which in turn
will decrease $N_{\rm eff}$.
But if flavor oscillations take place at temperatures
close to neutrino decoupling this would not hold and an extra contribution of neutrinos to
radiation is expected with respect to the value in Eq.\ (\ref{deltan_xinu}), as emphasized in
\cite{Pastor09} and shown in Fig.\ \ref{fig:asym}, where the
$N_{\rm eff}$ isocontours for non-zero mixing  are compared with those obtained from
the frozen neutrino distributions taking into account the effect of flavor oscillations
\cite{Mangano11}. One can see that oscillations efficiently reduce $N_{\rm eff}$
for neutrino asymmetries with respect to the initial values from Eq.\ (\ref{deltan_xinu}).

The evolution of the neutrino and antineutrino distribution functions
with non-zero initial asymmetries, from $T=10$ MeV until BBN,
has been calculated in \cite{Pastor09,Mangano11}. Here we
use the final numerical results for these spectra
in a range of values for $\eta_{\nu_e}^{\rm in}$ and $\eta_{\nu}$
as an input for our analysis, described in the next Section.
Note that an analysis in terms of the degeneracy parameters $\xi_\alpha$
as done for instance in \cite{Shiraishi09} is no longer possible.
We adopt the best fit values for the neutrino oscillation
parameters quoted in \cite{Schwetz11}, assuming a normal hierarchy of the
neutrino masses, except for the mixing angle $\Th$, for which we will
adopt two distinct values: $\Th=0$ and $\sin^2\Th=0.04$. The latter
is close to the upper limit placed by the Daya Bay \cite{dayabay12}
and RENO \cite{collaboration:2012nd} experiments
on this mixing angle (with a best-fit value of $\sin^2\Th=0.024$ and  
$\sin^2\Th=0.029$, respectively), and is used as an example to understand
the cosmological implications of a nonzero $\Th$. Moreover, since the
flavor asymmetries equilibrate for large values of this mixing angle,
the cosmological effects are similar for  $\sin^2\Th \gtrsim 0.02$, as in the case of an
inverted hierarchy for a broad range of $\Th$ values  (see,
for instance, Fig.~4 of Ref.~\cite{Mangano12}). As for the case $\Th=0$, though it seems presently disfavoured with a high statistical significance after the Daya Bay and RENO results, we have decided to include it for comparison.


\section{Cosmological constraints on neutrino parameters} \label{sec:theta13}

Having set the basic framework for the calculation
of the neutrino distribution functions in
the presence of asymmetries and for different 
$\Th$, we can now proceed to investigate its cosmological effects.

In order to constrain the values of the cosmological neutrino asymmetries, we compare our results to 
the observational data. In particular, we use a modified version of the CAMB code\footnote{\tt http://camb.info/} \cite{Lewis99}
to evolve the cosmological perturbations and obtain the CMB and matter power spectra in the presence of non-zero neutrino asymmetries in the neutrino distribution functions. We checked that the spectra computed by our modified CAMB version are consistent up to high accuracy with those obtained with CLASS \cite{Blas:2011rf}, that incorporates the models considered here in its public version. This version of CAMB is interfaced with the Markov chain Monte Carlo package
CosmoMC\footnote{\tt http://cosmologist.info/cosmomc/} \cite{Lewis02} that we use to sample the parameter 
space and obtain the posterior distributions for the parameters of interest. 

We derive our
constraints in the framework of a flat $\Lambda$CDM model with the three standard model neutrinos and purely adiabatic initial conditions. The parameters we use are described in Table \ref{tab:parameters} as well as the range of the flat priors used. As 
can be seen, six of them are the standard $\Lambda$CDM cosmological
parameters, and we add to those three new parameters, namely the mass of the lightest
neutrino mass eigenstate $m_1$ (the other two masses are calculated using the 
best fit for $\Delta m_{21}^2$ and $\Delta m_{31}^2$ obtained in \cite{Schwetz11}, assuming normal hierarchy) 
and the two neutrino asymmetries we mentioned
earlier, $\eta_{\nu_e}^{\ini}$ and $\eta_{\nu}$. The values of the effective degeneracy parameters $\xi_\alpha$ after BBN\footnote{The neutrino distribution functions can be parameterized by Fermi-Dirac-like functions with an effective $\xi_\alpha$ and  temperature $T_{\alpha}$ \cite{Mangano11},  
which are related to the first two moments of the distribution, the number density and energy density.}, needed by CAMB,
are pre-calculated as a function of the asymmetries (following the method described in the previous section) over a grid in $(\eta_{\nu_e}^{\ini},\,\eta_{\nu})$ and stored on a table, used for interpolation during the Monte Carlo run.

A comment on the parameterization is in order. It is a standard practice in cosmological analyses to parameterize the neutrino masses via
$\Omega_{\nu}h^{2}$ or equivalently $f_{\nu}\equiv\Omega_{\nu}/\Omega_{dm}$, and from that 
(assuming that neutrinos decoupled at equilibrium) derive the sum of neutrino masses, which are taken to be degenerate. 
The presence of lepton asymmetries dramatically changes this simple scheme. Now the neutrino number density is 
a complicated function of the $\eta$'s obtained from a non-equilibrium distribution function.
When $f_{\nu}$ is used, any effect related to the way in which the total neutrino density is shared among 
the different mass eigenstates is completely lost. 
In that sense, the parameterization used in this paper looks more physically 
motivated since energy densities of neutrinos are 
constructed from two fundamental quantities, namely their phase space distributions and their masses.

The most basic dataset that we consider only consists of the WMAP 7-year temperature and polarization anisotropy data. We will refer to it simply as ``WMAP''.  The likelihood is computed using the the WMAP likelihood code publicly available at the LAMBDA website\footnote{\texttt{http://lambda.gsfc.nasa.gov/}}. We marginalize over the amplitude of the Sunyaev-Zel'dovich signal.

In addition to the WMAP data, we also include the BBN measurement of the $^4$He mass fraction $Y_p$ from the data
collection analysis done in \cite{Iocco09}, in the form of a Gaussian prior
\be \label{eq:yp}
Y_p = 0.250 \pm 0.003 \ \ \ (1\sigma)\, . 
\ee
Indeed, some authors have recently reported a larger central value, $Y_p \sim 0.257$ \cite{Izotov:2010ca,Aver:2010wq,Aver:2010wd}, with quite different uncertainty determinations. In \cite{Aver:2011bw} using a Markov chain Monte Carlo technique already exploited in \cite{Aver:2010wd}, the primordial value of $^4$He decreased again to $Y_p = 0.2534 \pm 0.0083$, which is compatible at 1$\sigma$ with 
(\ref{eq:yp}). We will not use these results in our analysis, but we will comment on their possible impact in the following. We also note that in \cite{Mangano:2011ar} a robust upper bound $Y_p <0.2631$ (95 \% C.L.) has been derived based on very weak
assumptions on the astrophysical determination of $^4$He abundance, namely that the minimum
effect of star processing is to keep constant the helium content of a low-metallicity gas, rather than increase it, as expected. As we will show, the measurement of $Y_p$ currently dominates the constraints on the asymmetries: if we were to conservatively allow for larger uncertainties on that measurement, like for example those reported in \cite{Aver:2011bw}, our constraints from present data would correspondingly be weakened. 
Moreover, we decided not to use the Deuterium measurements since at the moment they are not competitive with Helium for constraining the asymmetries (see, {\it e.g.}, Fig.~6 of Ref.~\cite{Mangano11}), although there are recent claims that they could place strong constraints on $N_{\eff}$ at the level of $\Delta N_{\eff} \simeq \pm 0.5$ \cite{Pettini12}. This is a very interesting perspective but at the moment, Deuterium measurements in different QSO absorption line systems show a significant dispersion, much larger than the quoted errors.

The dataset that uses both 
WMAP 7-year data and the determination of  the primordial abundance of Helium as in (\ref{eq:yp}) will be referred to 
as ``WMAP$+$He''.
Measurements of $Y_p$ represent the best ``leptometer'' currently available, in the sense that they place the most 
stringent constraints on lepton asymmetries for a given baryonic 
density \cite{Serpico05}. The $^4$He mass fraction depends on the baryonic density, the electron neutrino 
degeneracy parameter and the effective number of neutrino families. Thus, in order
to consistently implement the above determination of $Y_p$ in our Monte Carlo analysis, we compute $\Delta N_{\eff}$ and $\xi_e$
coming from the distribution functions 
calculated with the asymmetries (as explained in the previous section) and store them on a table. During the CosmoMC run, we use this table
to obtain by interpolation the values $\Delta N_{\eff}$ and $\xi_e$ corresponding to given values of the asymmetries (which are the parameters actually used in the Monte Carlo), and finally to obtain $Y_p$ as a function of
$\Delta N_{\eff}$, $\xi_e$ and $\Omega_bh^2$. Notice that this approach
is slightly less precise than the one used in Refs. \cite{Mangano11, Mangano12},
where a full BBN analysis was performed,
but this approximation should suffice for our purposes, especially taking into
account that we will be comparing BBN limits on the asymmetries with
the ones placed by other cosmological data, that as we shall see are far less constraining. In any case, we have checked that the agreement between the interpolation scheme and the full BBN analysis is at the percent level.

We derive our constraints from parallel chains generated using the 
Metropolis-Hastings algorithm. For a subset of the models, we have 
also generated chains using the slice sampling method, in order to 
test the robustness of our results against a change in the algorithm. We 
use the Gelman and Rubin $R$ parameter to evaluate the convergence of 
the chains, demanding that $R-1 < 0.03$. The one- 
and two-dimensional posteriors are derived by marginalizing over the 
other parameters.

\begin{figure*}[htpb]
\includegraphics[scale=0.45]{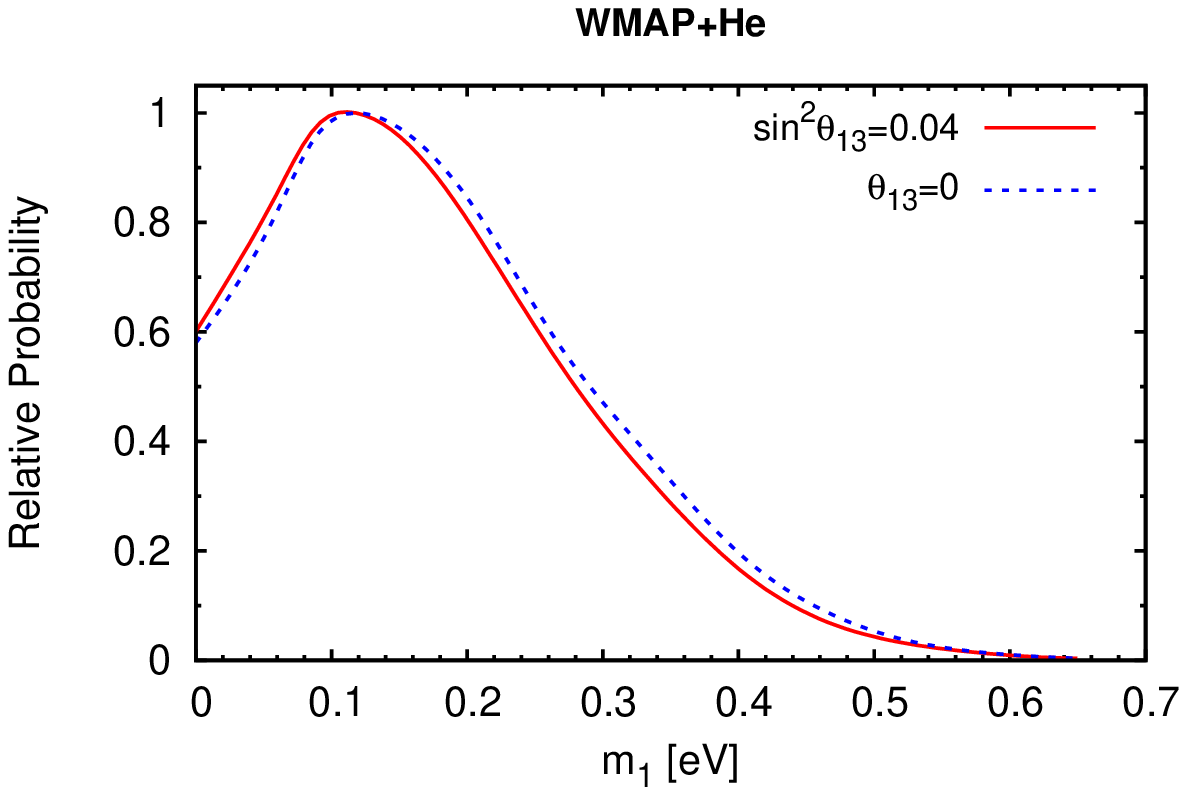}
\includegraphics[scale=0.45]{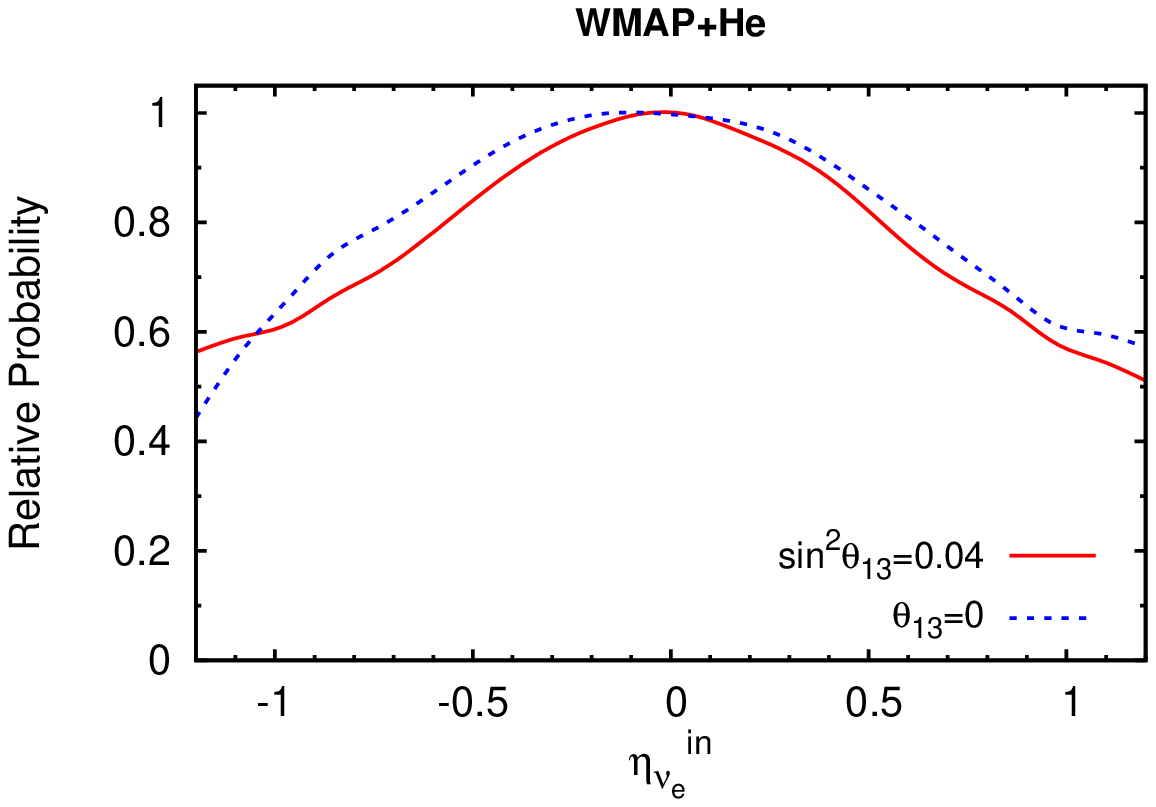}
\includegraphics[scale=0.45]{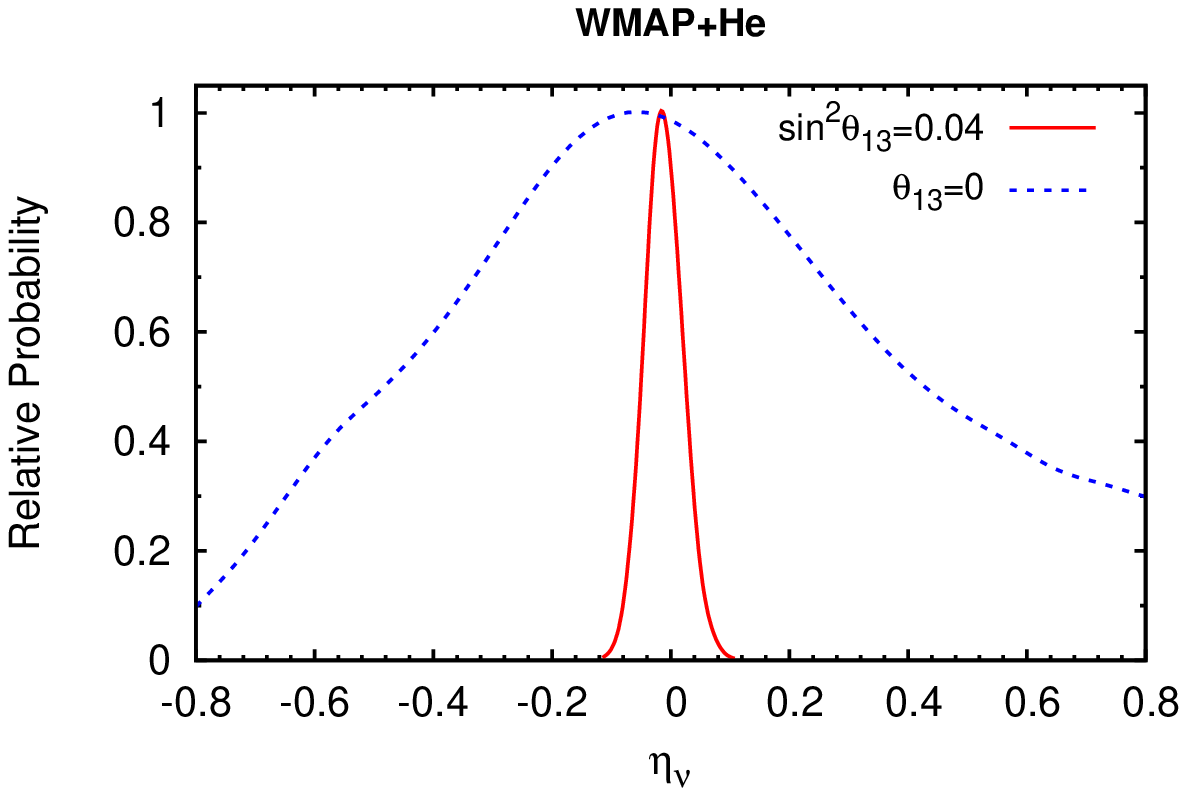}
\caption{ One-dimensional posterior probability density
for $m_1$, $\eta_{\nu_e}^{\ini}$, and  $\eta_{\nu}$ for the WMAP+He dataset. }
\label{fig:1D}
\end{figure*}

Our results for the cosmological and neutrino parameters from the analysis are shown in
Table \ref{tab:results}, while Fig.~\ref{fig:1D} shows the marginalized 
one-dimensional probability 
distributions for the lightest neutrino mass, 
the initial electron-neutrino asymmetry, and the 
total asymmetry, for the different 
values of $\Th$. Notice that the posterior for $\eta_{\nu_e}^{\ini}$ (middle panel) is still quite large at the edges of the prior range. This happens also for both the $\eta_{\nu_e}^{\ini}$ and $\eta_\nu$ posteriors obtained using only the WMAP data (not shown in the figure). Since the priors on these parameters do not represent a real physical constraint (as in the case $m_\nu>0$), but just a choice of the range to explore, we refrain from quoting 95\% credible intervals in these cases, as in order to do this one would need knowledge of the posterior in all the region where it significantly differs from zero. However, it is certain that the \emph{actual} 95\% C.I. includes the one that one would obtain using just part of the posterior (as long as this contains the peak of the distribution). If we do this, we obtain constraints that are anyway much worse than those from BBN. Finally, we also stress that if a larger experimental determination of $Y_p$ or measurements with larger uncertainities were used, as those reported in \cite{Izotov:2010ca,Aver:2010wq,Aver:2010wd}, BBN would show a preference for larger values of $N_{\eff}$ as well.

\begin{table*}[htpb]
\caption{ 95\% C.L.~constraints on cosmological parameters for the WMAP and WMAP+He datasets. }
\label{tab:results}
\setlength{\tabcolsep}{20pt}
\begin{tabular}{lcccc}
\hline \hline
Parameter & WMAP & \  &  WMAP+He & \ \\
$ \ $ & $\sin^2 \theta_{13} = 0$ & $\sin^2 \theta_{13} = 0.04$ & 
$\sin^2 \theta_{13} = 0$ & $\sin^2 \theta_{13} = 0.04$\\ 
\hline
100 $\Omega_b h^2$ & $2.20^{+0.14}_{-0.12}$ & $2.20^{+0.13}_{-0.12}$ & 
$2.20\pm 0.12$ & $2.20 \pm 0.12$ \\
$\Omega_{dm} h^2$ & $0.118 \pm 0.016$ & $0.117^{+0.017}_{-0.016}$ & 
$0.119 \pm 0.017 $ & $0.117 \pm 0.016$\\
$\tau$ & $0.085^{+0.029}_{-0.026}$  & $0.085^{+0.030}_{-0.027}$  &
$0.085^{+0.030}_{-0.027}$   & $0.085^{+0.029}_{-0.027}$  \\
$100 \theta_s $ & $1.0387 \pm 0.0063 $ & $1.0389^{+0.0069}_{-0.0063}$ &
$1.0381^{+0.054}_{-0.053}$ & $1.0387^{+0.0053}_{-0.0054}$ \\
$n_s$ & $0.953 \pm 0.032$ & $0.953^{+0.032}_{-0.033}$ & 
$0.955^{+0.034}_{-0.035}$ & $0.952^{+0.031}_{-0.032}$\\
$\log\left[ 10^{10} A_{s} \right]$ & $3.064^{+0.080}_{-0.082}$ & $3.062^{+0.080}_{-0.079}$ & 
$3.068^{+0.081}_{-0.078}$ & $3.062^{+0.073}_{-0.075}$ \\
\hline
$m_1$ (eV) &  $\le 0.39$ & $\le 0.38$ &  
$\le 0.38$ & $\le 0.38$ \\
$\eta_{\nu_e}^{\ini}$ & -- \footnote{The 95\% confidence region is not well-defined in these cases because the posterior does not vanish at the end of the prior range (see e.g. the middle panel of Fig. \ref{fig:1D}). See discussion in the text.} & -- $^\mathrm{a}$
& -- $^\mathrm{a}$ & -- $^\mathrm{a}$
\\
$\eta_{\nu}$ & -- $^\mathrm{a}$ & -- $^\mathrm{a}$
& $[-0.64;0.72]$ & $[-0.071;0.054]$\\
\hline
$h$ & $0.652^{+0.084}_{-0.083}$ & $0.653^{+0.081}_{-0.082}$ 
& $0.656^{+0.084}_{-0.081}$ & $0.650^{+0.078}_{-0.081}$ \\
$\Delta N_{\eff}$ & $\le 0.32$ & $\le 0.16$ 
& $\le 0.43$ & $\le 0.03$\\
\hline
\end{tabular}
\setlength{\tabcolsep}{6pt}
\end{table*}

\begin{figure}[htpb]
\begin{center}
\includegraphics[scale=0.6]{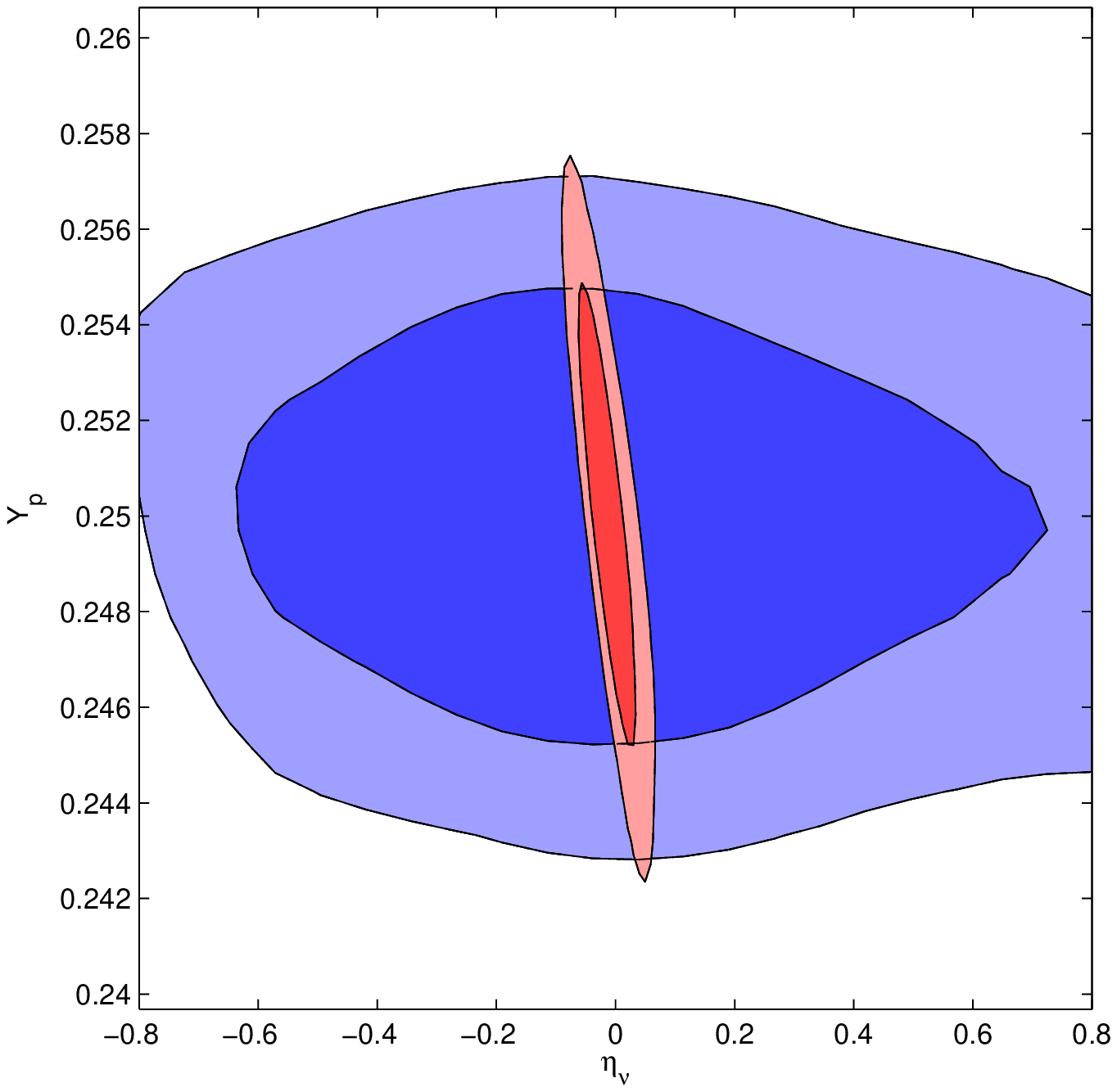}
\caption{
68\% and 95\% confidence regions in total neutrino asymmetry $\eta_{\nu}$ vs.~the primordial abundance
of Helium $Y_p$ plane for $\Th=0$ (blue) and $\sin^2\Th=0.04$ (red), from the analysis of the WMAP+He dataset. Notice the much stronger 
constraint for the nonzero mixing angle due to the faster equilibration of flavor asymmetries.}
\label{fig:eta_Y}
\end{center}
\end{figure}

Concerning the neutrino asymmetries, shown in the middle and right panels of Fig.~\ref{fig:1D},
we notice that  while the initial flavor asymmetries remain highly unconstrained by current
data, the total asymmetry constraint improves significantly for 
$\Th \neq 0$. This result agrees with previous results from BBN-only studies \cite{Mangano11, Mangano12}, 
and it is a result of the equilibration of flavor asymmetries when 
$\Th$ is large (see, {\it e.g.}, Fig.~5 of Ref.~\cite{Mangano11}). When the 
flavors equilibrate in the presence of a nonzero mixing angle ($\sin^2\Th=0.04$ in
our example) the total asymmetry 
is distributed almost equally among the different flavors, leading to
a final asymmetry $\eta_{\nu_e}^{\fin} \approx \eta_{\nu_x}^{\fin} \approx \eta_{\nu} / 3$ (where $x=\mu,\tau$).
Hence, the fact that the BBN prior requires $\eta_{\nu_e}^{\fin} \approx 0$ for the correct abundance of
primordial Helium (see Fig.~\ref{fig:eta_Y}) leads to a strong constraint on the constant total asymmetry, 
$-0.071 \leq \eta_{\nu} \leq 0.054$ (95$\%$ C.L.).

\begin{figure}[htbp] 
\includegraphics[scale=0.6]{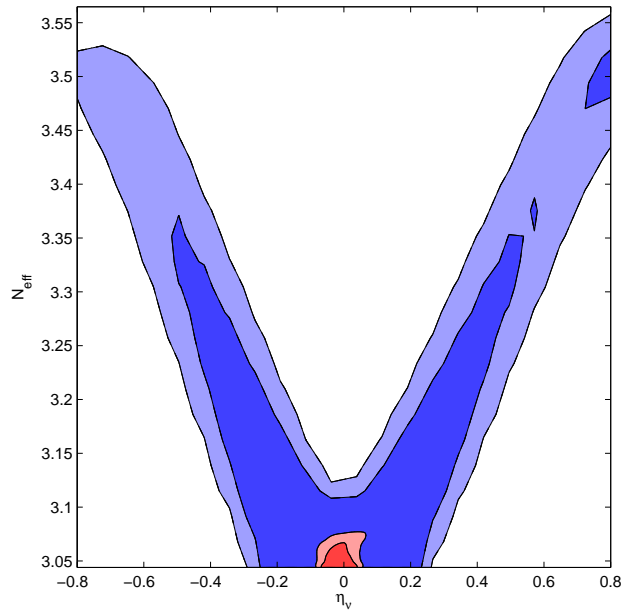}
\caption{Two-dimensional 68\%  and 95\% confidence regions in the $(\eta_\nu,\,N_\mathrm{eff}$) plane from the analysis of the WMAP+He dataset, for $\Th=0$ (blue) and $\sin^2\Th=0.04$ (red). Even for
zero $\Th$ the data seem to favor $N_{\eff}$ around the
standard value $N_{\eff}=3.046$.}
\label{fig:neff2D}
\end{figure}

On the other hand, since the constraints come most from the distortion in the electron neutrino distribution
function,
when $\Th=0$ (and therefore there is less mixing) the direct relation between  $\eta_{\nu_e}^{\fin}$ and
$\eta_{\nu}$ is lost. In this case, the total asymmetry could still be large, even if 
the final electron neutrino asymmetry is small, as significantly asymmetries
can still be stored on the other two flavors, leading to a constraint
an order of magnitude weaker than the previous case, $-0.64 \leq \eta_{\nu} \leq 0.72$ (95\% C.L.).
As expected, this is reflected on the allowed ranges for $\Delta N_{\eff}$, as shown in Fig.\ 
\ref{fig:neff2D}: while for $\Th = 0$ the $\Delta N_{\eff} \simeq 0.5$ are still allowed
by the data, nonzero values of this mixing angle reduce the allowed region in the parameter
space by approximately an order of magnitude in both $\Delta N_{\eff}$ and $\eta_{\nu}$.

\begin{figure}[htbp] 
\includegraphics[scale=0.7]{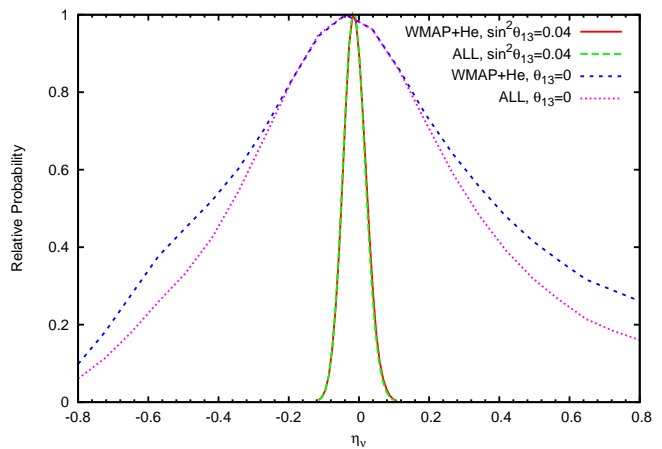}
\caption{One-dimensional posterior probability density
for $\eta_{\nu}$ comparing the WMAP+He and the ALL datasets.
As mentioned in the text, the constraints on the total asymmetry 
do not improve significantly with the inclusion of other cosmological datasets, 
as they are mainly driven by
the determination of the primordial Helium abundance.}
\label{fig:bbn_all}
\end{figure}

We confirmed in our analysis that
the constraints on the asymmetry are 
largely dominated by the BBN prior at present. This is shown in 
Fig.~\ref{fig:bbn_all}, where we compare the results of our analysis
with a more complete dataset (which we refer to as ALL) 
that includes distance measurements of SNIa from the SDSS compilation \cite{Kessler:2009ys}
and the HST determination of the Hubble constant $H_0$ \cite{Riess:2009pu}, as well as
data on the power spectrum of the matter density field, as reconstructed from a sample of 
Luminous Red Galaxies of the SDSS Seventh Data Release \cite{Reid:2009xm}.
This is due to the fact that other
cosmological data constrain the asymmetries via their
effect on increasing $N_{\eff}$, and currently the errors
on the measurement of the effective number of neutrinos
\cite{Komatsu11,act11,spt11,spt12} are significantly weaker 
than our prior
on $Y_p$, eq.~(\ref{eq:yp})\footnote{On the other hand, these other cosmological data sets have an impact on other parameters like e.g. the neutrino mass. But since in this work we are primarily interested in bounding the asymmetries, we prefer to stick to the robust WMAP+He data set. In that way, our results are not contaminated by possible systematic uncertainties in the other data. Actually, the inclusion of all external datasets (in particular, of SNIa together with $H_0$) reveals a conflict between them, leading to a bimodal posterior probability for $\Omega_{dm}h^2$ and to a preference for $m_1>0$ at 95\% C.L.}. The fact that bounds on leptonic asymmetries are dominated by the BBN
prior (i.e. by $^4$He data) is also confirmed by the similarity of our
bounds on ($\eta_\nu$, $\eta_{\nu_e}^{\rm in}$) with those of \cite{Mangano12}. Note
that the limits reported in [35] sound weaker, because they are
frequentist bounds obtained by cutting the parameter probability at
$\Delta \chi^2=6.18$, i.e. they represent 95\% bounds on joint
two-dimensional parameter probabilities (in the Gaussian
approximation). The one-dimensional 95\% confidence limits,
corresponding to $\Delta \chi^2=4$, are smaller and very close to the
results of the present paper. We also checked that using our codes and data sets, we obtain very similar results when switching from Bayesian to
frequentist confidence limits.

We conclude this section noting that the current
constraints on the sum of neutrino masses are robust under
a scenario with lepton asymmetries, as those extra 
degrees-of-freedom  do not correlate
with the neutrino mass. On the other hand, to
go beyond the BBN limits on the asymmetries more
precise measurements of $N_{\eff}$ are clearly needed,
and in the next section we forecast the results
that could be achievable with such an improvement
using COrE as an example of future CMB experiments.

\section{Forecast} \label{sec:forecast}

Given that the current constraints on the lepton asymmetries
are dominated by their effect on the primordial production of
light elements, one can ask whether future cosmological
experiments can improve over the current limits imposed
by BBN. With that goal in mind, we take as an example a 
proposed CMB experiment, COrE (Cosmic Origins Explorer) \cite{core11},
designed to detect the primordial gravitational waves and 
measure the CMB gravitational lensing deflection power spectrum 
on all linear scales to the cosmic variance limit. The latter is
of special interest for this work, as the CMB lensing is expected to 
probe with high sensitivity the absolute neutrino masses and $N_{\eff}$ 
\cite{Lesgourgues05}.

We used the package FuturCMB\footnote{\tt http://lpsc.in2p3.fr/perotto/} in combination with CAMB and CosmoMC for producing mock CMB data, and fit it with a likelihood based on the potential sensitivity of COrE. We include, also in this case, the information coming from present measurements of the Helium fraction, encoded in the Gaussian prior (\ref{eq:yp}).  We consider five of COrE's frequency channels, ranging from 105 to 225 GHz, with the specifications given in \cite{core11} and reported for convenience in Table \ref{tab:corespec}, and assume an observed fraction $f_\mathrm{sky} = 0.65$. We do not consider other channels as they are likely to be foreground dominated. We take a maximum multipole $\ell_\mathrm{max}=2500$. In our analysis, we have assumed that the uncertainties associated to the beam and foregrounds have been properly modeled and removed, so that we can only consider the statistical uncertainties. Those are optimistic assumptions, as under realistic conditions 
systematic uncertainties will certainly play an important role. In that sense, our results represent an illustration of what future CMB experiments could ideally achieve.

\begin{table}[!htb]
\begin{center}
\setlength{\tabcolsep}{7pt}
\begin{tabular}{cccc}
\hline
\hline
Frequency [GHz]& $\theta_\mathrm{fwhm}$  [arcmin] & $\sigma_T$ [$\mu$K]&  $\sigma_P $  [$\mu$K] \\  
\hline
105   &  10.0   &  0.268    &  0.463  \\
135   &    7.8   &  0.337    &  0.583  \\
165   &    6.4   &  0.417    &  0.720  \\
195   &    5.4   &  0.487    &  0.841  \\
225   &    4.7   &  0.562    &  0.972  \\
\hline
\end{tabular}
\caption{Experimental specifications for COrE \cite{core11}. For each channel, we list the channel frequency in GHz, the FWHM in arcminutes, the temperature ($\sigma_T$) and polarization ($\sigma_P$) noise per pixel in $\mu$K.}
\label{tab:corespec}
\end{center}
\end{table}

%

We use CMB lensing information in the way described in \cite{Perotto06}, assuming that the CMB lensing potential spectrum will be extracted from COrE maps with a quadratic estimator technique.

For the forecast we adopt the fiducial values for the cosmological 
parameters shown in Table \ref{tab:fiducial} for both cases of
$\theta_{13}$ discussed previously. The two sets of fiducial values correspond to the 
best-fit models of the WMAP+He dataset for the two values of $\Th$. In the case of the neutrino mass, since the likelihood is 
essentially flat between $0$ and $0.2$ eV, we have chosen to take $m_1 = 0.02$ eV. This is below the expected sensitivity of COrE 
and should thus be essentialy equivalent to the case where the  lightest neutrino is massless. 

\begin{table}[ht] 
\caption{Fiducial values for the cosmological parameters for the COrE forecast.}
\setlength{\tabcolsep}{12pt}
\begin{tabular}{lcc} 
\hline \hline 
Parameter & Fiducial Value & Fiducial Value \\ 
$\ $ & ($\sin^2 \theta_{13} = 0$) & ($\sin^2 \theta_{13} = 0.04$) \\ \hline
$\Omega_b h^2$ & $0.0218$ & $0.0224$ \\ 
$\Omega_{dm} h^2$ & $0.121$ & $0.118$ \\ 
$\tau$ & $0.0873$ & $0.0865$ \\
$h$ & $0.709$ & $0.705$\\
$n_s$ & $0.978$ & $0.968$ \\
$\log\left[ 10^{10} A_{s} \right]$ & $3.12$ & $3.08$ \\
\hline 
$m_1$ (eV) &  $0.02$ & $0.02$ \\
$\eta_{\nu_e}^{\ini}$ & $0$ & $0$ \\
$\eta_{\nu}$ & $0$ & $0$ \\
\hline
\end{tabular} \label{tab:fiducial}
\setlength{\tabcolsep}{6pt}
\end{table}

\begin{figure*}[htbp]
\begin{center}
\includegraphics[scale=0.46]{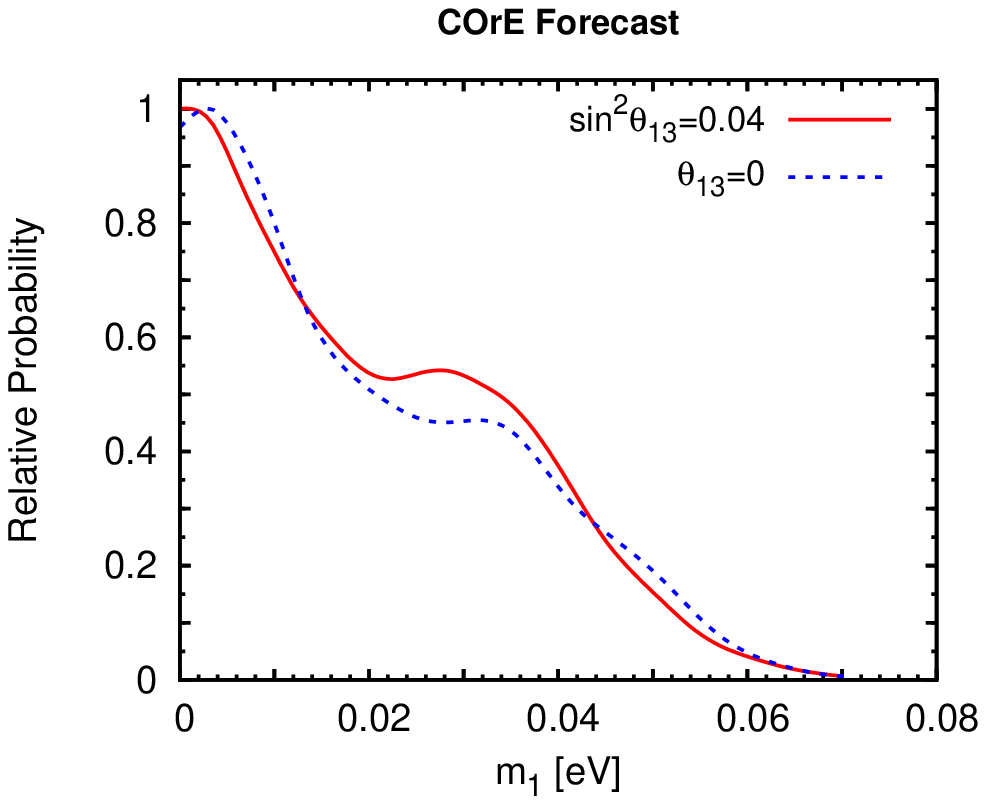}
\includegraphics[scale=0.46]{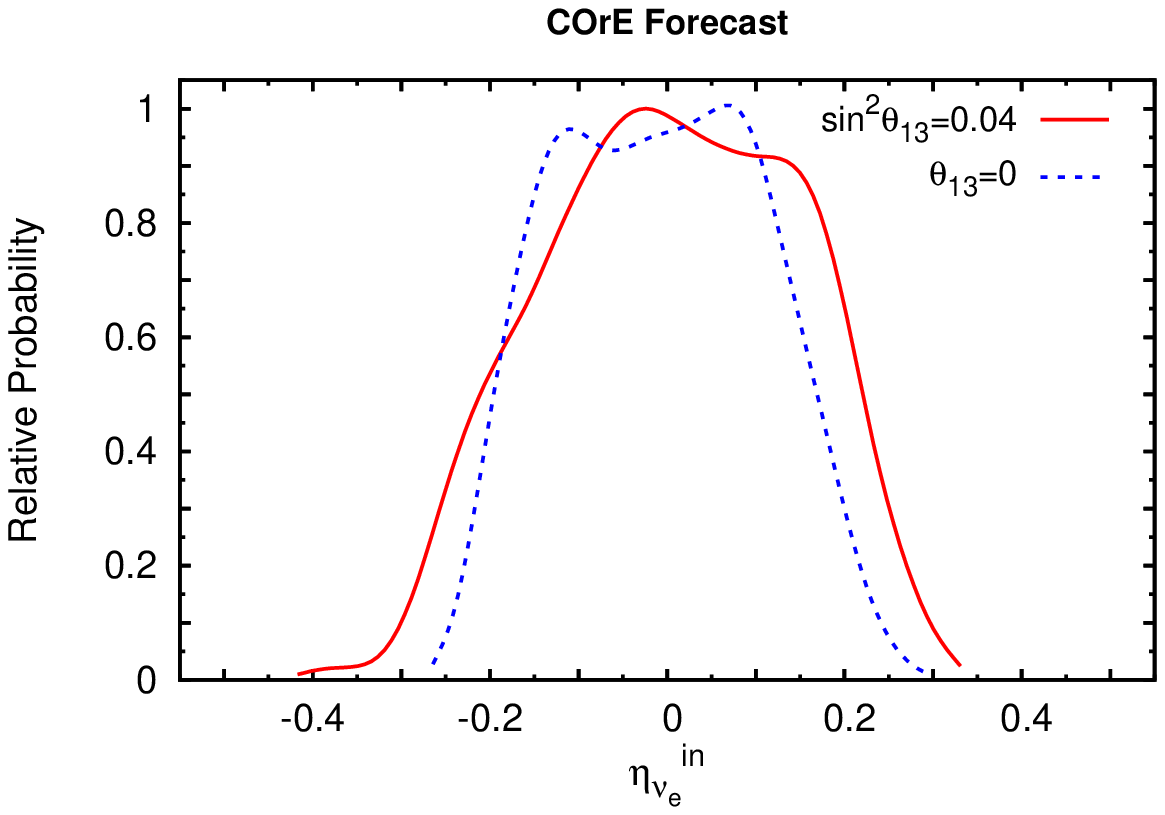}
\includegraphics[scale=0.46]{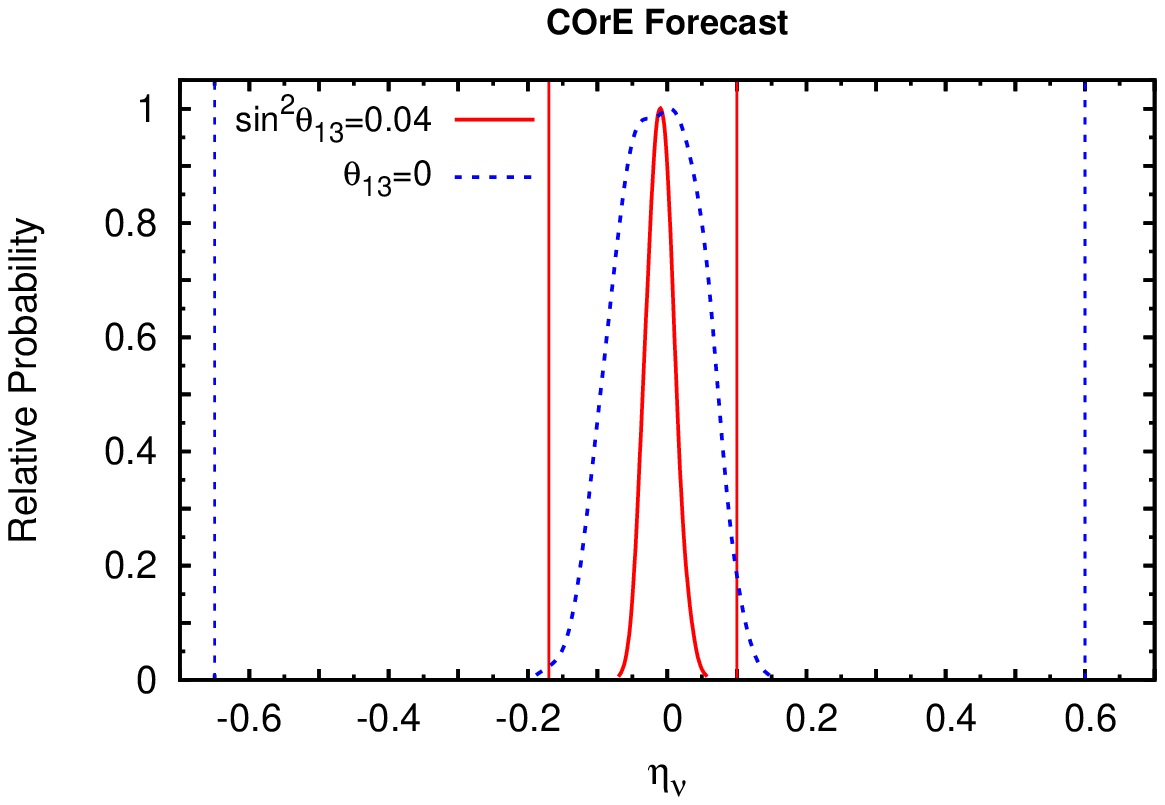}
\caption{One-dimensional probability distribution 
function for $m_1$ and  $\eta_{\nu}$ for COrE
forecast. The middle panel shows that an experiment like COrE could start constrain 
the initial electron neutrino asymmetry. The vertical lines on the right panel show the current
95$\%$ C.L. limits obtained in the previous section. 
The errors on the asymmetries are
improved by approximately a factor 6.6 or 1.6 for $\Th=0$ and $\sin^2 \theta_{13} = 0.04$, respectively, compared to the 
results shown in Fig.~\ref{fig:1D}.}
\label{fig:forecast}
\end{center}
\end{figure*}

The sensitivities on the neutrino parameters for COrE are shown in 
Fig.~\ref{fig:forecast} for the two values of $\Th$. As expected for the
sum of the neutrino masses, the constraints are significantly better
than the current ones, and could in principle start probing the
minimal values guaranteed by flavor oscillations
\cite{Lesgourgues05}. Note that our forecast 
error for $m_1$ differs slightly from the one presented in \cite{core11}, 
most probably because the forecasts in this reference are based on the Fisher matrix approximation.
But our main goal in this section is to discuss how
COrE observations will help improving the
limits on the asymmetries discussed previously, that are basically dominated by the available measurements of the $^4$He abudance.
The right panel of Fig.~\ref{fig:forecast}
shows the forecasted posterior probability distribution 
for $\eta_{\nu}$, and the marginalized
constraints for it are listed in Table \ref{tab:errors} for both values
of $\Th$; in particular, the vertical lines of the right panel
show the 95\% C.L.~limits obtained from the full BBN analysis of Ref. \cite{Mangano12}. 
Comparing
the values from Tables \ref{tab:results} and \ref{tab:errors}
one can see that an experiment like COrE would improve current 95\% limits on 
the total leptonic asymmetry by nearly a factor 6.6 ($\Th=0$) and 1.6 ($\sin^2{\Th}=0.04$), 
competitive over the constraints from $^4$He abundance only. It should be noted that the 
error bars on the primordial abundances are very difficult to be reduced due
to systematic errors on astrophysical measurements \cite{Iocco09},
and therefore it is feasible that CMB experiments will
be an important tool in the future to improve the constraints on 
the asymmetries. Notice however that, since the CMB is insensitive to the sign of the $\eta$'s, BBN measurements will still be needed in order to break this degeneracy.

\begin{table}[t] 
\caption{95\% confidence intervals for the neutrino parameters with COrE. }
\setlength{\tabcolsep}{14pt}
\begin{tabular}{lcc} 
\hline \hline 
Parameter & $\sin^2 \theta_{13} = 0$
&  $\sin^2 \theta_{13} = 0.04$ \\ \hline
$ m_1$ (eV) &  $< 0.049$ & $ < 0.048$ \\
$ \eta_{\nu_e}^{\ini}$ & $[-0.20;0.20]$  & $[-0.25; 0.24]$\\
$ \eta_{\nu}$ & $[-0.12;0.09]$  & $[-0.048; 0.030]$\\
\hline
\end{tabular} \label{tab:errors}
\setlength{\tabcolsep}{6pt}
\end{table}
\begin{figure*}[htbp]
\begin{center}
\includegraphics[scale=0.67]{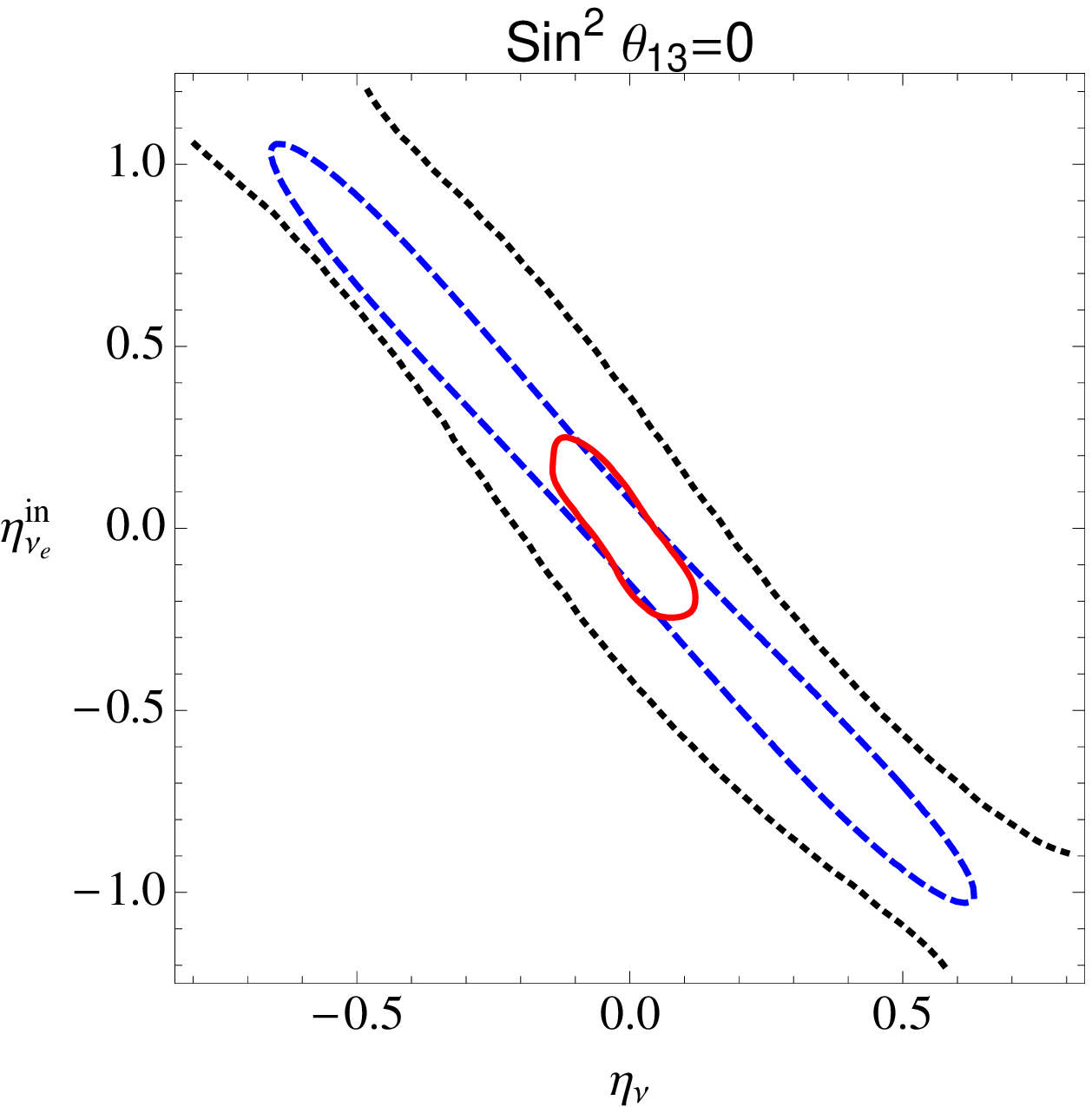}
\includegraphics[scale=0.67]{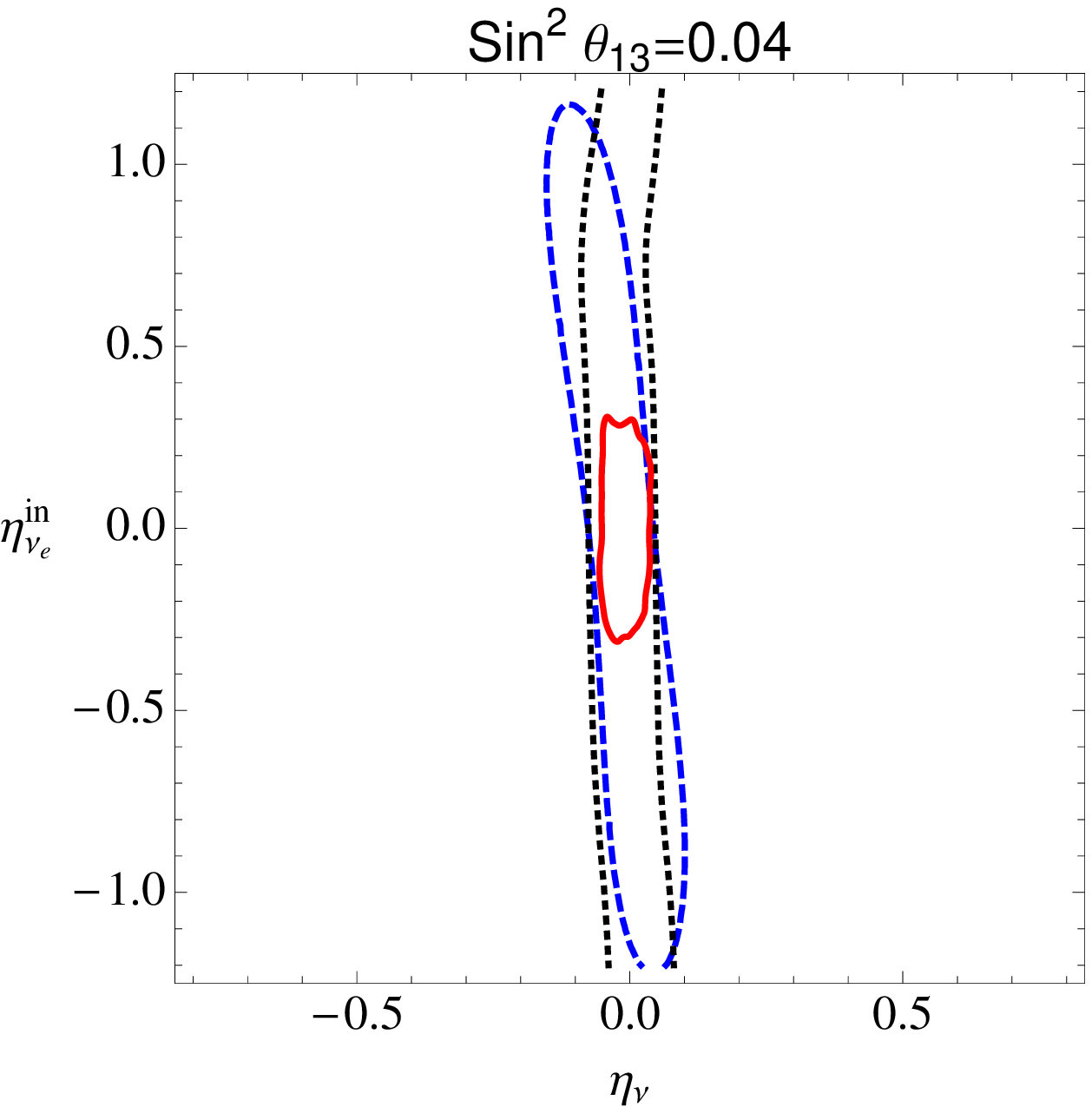}
\caption{The 95$\%$ C.L.~ contours on the $\eta_{\nu}$ vs. $\eta_{\nu_e}^{\ini}$ plane 
from our analysis with current data (WMAP+He dataset, black dotted) compared to the results of the BBN analysis  
of Ref. \cite{Mangano12}
(blue dashed) and with the COrE forecast (red solid).} 
\label{fig:comparison}
\end{center}
\end{figure*}

Finally, in Fig.~\ref{fig:comparison} we show the COrE sensitivity on
the asymmetries in the plane $\eta_{\nu}$ vs. $\eta_{\nu_e}^{\ini}$ compared to 
the constraints of Sec.~\ref{sec:theta13} obtained using current data  
and to the full BBN analysis of Ref.~\cite{Mangano12}. Notice that in the case $\Th=0$
 the constraints
of the previous section are quite less constraining than the ones coming
from the full BBN analysis because we are not using deuterium data, known to
be important to close the contours on the asymmetries plane, especially
for small values of $\Th$ \cite{Mangano11}. Moreover, future
CMB experiments have the potential to reduce the allowed region, 
dominating the errors in this analysis.

In summary, an experiment like COrE is capable of improving the constraints
on the lepton asymmetries by up to a factor 6.6 on
the total and/or flavor asymmetries depending on the
value of the mixing angle $\Th$. In addition to
that, such an experiment would also constrain other cosmological
parameters (in particular the sum of the neutrino masses)
with significant precision, providing yet another step towards
the goal of accurately measuring the properties of
the Universe.


\section{Conclusions} \label{sec:conc}

Understanding the physical processes that took place in the
early Universe is a crucial ingredient for deciphering the
physics at energies that cannot be currently probed in terrestrial
laboratories. In particular, since the origin of the matter-antimatter
is still an open question in cosmology, it is important to keep an
open mind for theories that predict large lepton
asymmetries. In that case, constraining total and flavor neutrino 
asymmetries using cosmological data 
is a way to test and constrain some 
of the possible particle physics scenarios at epochs 
earlier than the BBN.

For that, we initially used current cosmological
data to constrain not only the asymmetries, but also to understand
the robustness of the cosmological parameters (and the
limits on the sum of the neutrino masses) 
for two different values of the mixing angle
$\Th$ to account for the evidences of a nonzero value for this
angle.  Our results confirm the fact that at present
the limits on the cosmological lepton asymmetries 
are dominated by the abundance
of primordial elements generated during the BBN, in 
particular the abundance of $^4$He, currently the
most sensitive ``leptometer'' available. 

However, future CMB experiments might be able to
compete with BBN data in what concerns 
constraining lepton asymmetries, although BBN will always be needed in order
to get information on the sign of the  $\eta$'s.
We took as an example 
the future CMB mission COrE, proposed to measure
with unprecedent precision the lensing of CMB anisotropies,
and our results indicate that it has the potential to 
significantly improve
over current constraints while, at the same time placing limits 
on the sum of the neutrino masses that are of the
order of the neutrino mass differences.

Finally, we notice that for the values of $\Th$ 
measured by the Daya Bay and RENO experiments the limits on the
cosmological lepton asymmetries and on its associated
effective number of neutrinos are quite strong, so that lepton asymmetries cannot increase 
$N_{\eff}$ significantly above 3.4. 
Under those circumstances, if the cosmological data (other than
BBN) continues to push for large values of $N_\eff$, 
new pieces of physics such as sterile neutrinos  
will be necessary to explain that excess.

\section*{Acknowledgments}
We thank Srdjan Sarikas for providing some of the data used
in Figs.\ \ref{fig:asym} and \ref{fig:comparison}.
The work of AM was supported by the PRIN-INAF grant ``Astronomy probes
fundamental physics'' and
by the Italian Space Agency through the ASI contract Euclid- IC (I/031/10/0).
GM acknowledges support by
the {\it Instituto Nazionale di Fisica Nucleare} I.S. FA51 and the PRIN
2010 ``Fisica Astroparticellare: Neutrini ed Universo Primordiale'' of the
Italian {\it Ministero dell'Istruzione, Universit\`a e Ricerca}. 
The work of ML is supported by Ministero dell'Istruzione, dell'Università e 
della Ricerca (MIUR) through the PRIN grant ``Matter-antimatter asymmetry, 
Dark Matter and Dark Energy in the LHC era'' (contract number PRIN 2008NR3EBK-005). SP and UF were 
supported by the Spanish grants FPA2008-00319, FPA2011-22975 and Multidark CSD2009-00064
(MINECO) and PROMETEO/2009/091 (Generalitat Valenciana), and by the EC
contract UNILHC PITN-GA-2009-237920. UF acknowledges the support of the I3P-CSIC 
fellowship and of EPLANET. This research was also supported by a
Spanish-Italian MINECO-INFN agreement, refs.\ AIC10-D-000543 and AIC-D-2011-0689.
Finally, EC acknowledges the hospitality of CERN while working on 
this paper. 


\end{document}